\documentstyle[twoside]{article}

\catcode`\@=11
\long\def\@makefntext#1{
\protect\noindent \hbox to 3.2pt {\hskip-.9pt  
$^{{\eightrm\@thefnmark}}$\hfil}#1\hfill}		

\def\@makefnmark{\hbox to 0pt{$^{\@thefnmark}$\hss}}	
	
\def\ps@myheadings{\let\@mkboth\@gobbletwo
\def\@oddhead{\hbox{}
\rightmark\hfil\eightrm\thepage}   
\def\@oddfoot{}\def\@evenhead{\eightrm\thepage\hfil
\leftmark\hbox{}}\def\@evenfoot{}
\def\sectionmark##1{}\def\subsectionmark##1{}}

\oddsidemargin=\evensidemargin
\newcommand{\textlineskip}{\baselineskip=13pt}
\newcommand{\smalllineskip}{\baselineskip=10pt}

\def\eightcirc{
\begin{picture}(0,0)
\put(4.4,1.8){\circle{6.5}}
\end{picture}}
\def\eightcopyright{\eightcirc\kern2.7pt\hbox{\eightrm c}} 

\newcommand{\copyrightheading}[1]
	{\vspace*{-2.5cm}\smalllineskip{\flushleft
        {\footnotesize 
        Proc. Roy. Irish Acad. 47 A (October 1941) 53-54
        }\\
	 }}

\newcounter{itemlistc}
\newcounter{romanlistc}
\newcounter{alphlistc}
\newcounter{arabiclistc}

\def\@citex[#1]#2{\if@filesw\immediate\write\@auxout
	{\string\citation{#2}}\fi
\def\@citea{}\@cite{\@for\@citeb:=#2\do
	{\@citea\def\@citea{,}\@ifundefined
	{b@\@citeb}{{\bf ?}\@warning
	{Citation `\@citeb' on page \thepage \space undefined}}
	{\csname b@\@citeb\endcsname}}}{#1}}

\def\@refcitex[#1]#2{\if@filesw\immediate\write\@auxout
	{\string\citation{#2}}\fi
\def\@citea{}\@refcite{\@for\@citeb:=#2\do
	{\@citea\def\@citea{, }\@ifundefined
	{b@\@citeb}{{\bf ?}\@warning
	{Citation `\@citeb' on page \thepage \space undefined}}
	\hbox{\csname b@\@citeb\endcsname}}}{#1}}

\def\@refcite#1#2{{#1\if@tempswa\typeout
        {IJCGA warning: optional citation argument
	ignored: `#2'} \fi}}

\def\refcite{\@ifnextchar[{\@tempswatrue
	\@refcitex}{\@tempswafalse\@refcitex[]}}

\def\pmb#1{\setbox0=\hbox{#1}
	\kern-.025em\copy0\kern-\wd0
	\kern.05em\copy0\kern-\wd0
	\kern-.025em\raise.0433em\box0}

\def\fnt#1#2{\footnotetext{\kern-.3em
	{$^{\mbox{\scriptsize #1}}$}{#2}}}

\def\runninghead#1#2{\pagestyle{myheadings}
\markboth{{\protect\footnotesize\it{\quad #1}}\hfill}
{\hfill{\protect\footnotesize\it{#2\quad}}}}
\font\eightrm=cmr8

\def\qed{\hbox{${\vcenter{\vbox{			
   \hrule height 0.4pt\hbox{\vrule width 0.4pt height 6pt
   \kern5pt\vrule width 0.4pt}\hrule height 0.4pt}}}$}}

\begin{document}



\runninghead{Erwin Schroedinger
$\ldots$} {Erwin Schroedinger
$\ldots$}


\thispagestyle{empty}
\setcounter{page}{1}

\centerline{Proceedings Royal Irish Academy {\bf 47} A, 53-54 (1941)}

\vspace*{0.88truein}

\centerline{IV.}
\bigskip
\centerline{\bf THE FACTORIZATION OF THE HYPERGEOMETRIC EQUATION}
\vspace*{0.035truein}
\vspace*{0.37truein}
\centerline{\footnotesize by ERWIN SCHR\"ODINGER.}
\centerline{\footnotesize (From the Dublin Institute for Advanced Studies.)}
\vspace*{0.015truein}
\centerline{\footnotesize [Read 23 June. \hspace{1cm}
Published 29 December, 1941.]}
\vspace*{0.015truein}
\centerline{[\footnotesize{\it in LaTex by} Haret C. Rosu
(October 1999)]}
\baselineskip=10pt
\vspace*{10pt}
\vspace*{0.225truein}

\vspace*{0.21truein}


\textlineskip                  
\vspace*{12pt}                 

\vspace*{1pt}\textlineskip	
\vspace*{-0.5pt}
\noindent


\noindent





\noindent
As a sequel to investigations in factorizing ordinary homogeneous linear
differential equations of the second order\footnote{Proc. R.I.A. {\bf 46} A
(1940) 9; {\em ibid}, (1941) 183.},  I here indicate a quadruple of
factorizations of the hypergeometric equation, the one that determines
Gauss's function $F(\alpha, \beta, \gamma, x)$, of which most of the
functions occurring in physics are either special or limiting cases. The
equation reads
$$
x(1-x)y^{''}+[\gamma -(\alpha +\beta +1)x]y^{'}-\alpha \beta y=0~.
\eqno(1)
$$
In physical applications $x$ is usually restricted to
$$
0\leq x\leq 1~.
\eqno(2)
$$
If by
$$
\cos \theta =2x-1
\eqno(3)
$$
you introduce the independent variable $\theta$ (which by (2) would be
restricted to
$$
\pi \geq \theta \geq 0)~,
\eqno(4)
$$
you get
$$
\frac{d^2y}{d\theta ^2}+\frac{a\cos \theta +b}{\sin \theta}\frac{dy}{d\theta}
+cy~,
\eqno(5)
$$
with
$$
a=\alpha +\beta~, \qquad b=\alpha +\beta +1 -2\gamma~, \qquad c=-\alpha \beta~.
\eqno(6)
$$
If now you introduce the new {\em dependent} variable
$$
z=(\sin \theta)^{\frac{a}{2}}\left(
{\rm tan} \frac{\theta}{2}\right)^{\frac{b}{2}}y~.
\eqno(7)
$$
you obtain
$$
\frac{d^2z}{d\theta ^2}+\Big[c+\frac{a^2}{4}-\frac{2b(a-1)\cos \theta +a^2+b^2
-2a}{4\sin ^2\theta}\Big]z=0~.
\eqno(8)
$$
This is readily factorized thus
$$
\left(\frac{d}{d\theta}+\frac{C}{\sin \theta}+D {\rm cot} \theta \right)
\left(\frac{d}{d\theta}-\frac{C}{\sin \theta}-D {\rm cot} \theta\right)z+Bz=0~.
\eqno(9)
$$
Comparing the coefficients the following {\em four} alternatives are offered: -

\bigskip

\begin{tabular} {lll}
(1) $B=c$~, & $C=\frac{b}{2}$~, & $D=\frac{a}{2}$~.\\
(2) $B=c+a-1$~, & $C=-\frac{b}{2}$~, & $D=1-\frac{a}{2}~.
\qquad \quad
\qquad \qquad \qquad (10)$\\
(3) $B=c+\frac{a^2}{4}-\frac{(b+1)^2}{4}$~,
& $C=\frac{a-1}{2}$~, & $D=\frac{b+1}{2}$~.\\
(4) $B=c+\frac{a^2}{4}-\frac{(b-1)^2}{4}$~,
& $C=-\frac{a-1}{2}$~, & $D=-\frac{b+1}{2}$~.
\end{tabular}

\bigskip

It will be realized that it is the factorizations (3) and (4) which lend
themselves to the recurrent process described earlier. For they are obtained
from one another by reversing the order of the first order operators in (9)
and changing the value of $b$ by $\pm 2$, whilst $a$ and $c$ are unchanged.
(From (6) that means that $\gamma$ alone is changed, $\alpha$ and $\beta$
remaining constant.) If the particular problem is such as to warrant
$B\geq 0$, the recurrent process, in one or the other direction, must
lead to a function for which
$$
\left(\frac{d}{d\theta}-\frac{C}{\sin \theta}-D{\rm cot}\theta\right)z=0~,
$$
and
$$
B=0~.
$$
From this key-function the other solutions are obtained by repeated
application of the {\em other} operator.

The factorizations (9), (10) must not be regarded as {\em the} factorizations
of Gauss's equation. They belong to the particular {\em density}
$$
\sigma=(\sin \theta)^{a}\left({\rm tan} \frac{\theta}{2}\right)^{b}~,
$$
according to (7). There are bound to exist others belonging to other density
functions. They are {\em not} obtained just by a change of the dependent
variable
$$
\hat{z}=f(\theta)z~.
$$
This is a trivial transformation, which does not yield anything new.

\end{document}